\documentstyle[12pt]{article}
\begin{document}
\thispagestyle{empty}
\begin{center}
\LARGE \tt \bf {Non-Riemannian cosmic walls from gravitational collapse}
\end{center}
\vspace{2.5cm}
\begin{center} {\large By
L.C. Garcia de Andrade\footnote{Departamento de
F\'{\i}sica Te\'{o}rica - Instituto de F\'{\i}sica - UERJ
Rua S\~{a}o Fco. Xavier 524, Rio de Janeiro, RJ
Maracan\~{a}, CEP:20550-003 , Brasil.}}
\end{center}
\vspace{2cm}
\begin{abstract}
Two classes of metrics obtained from non-Riemannian gravitational collapse are presented.The first is the Taub planar symmetric exact solutions of Einstein-Cartan field equations of gravity describing torsion walls which are obtained from gravitational collapse of time dependent perturbation of Riemannian Taub symmetric solutions of General Relativity.The second is a modification of the Vilenkin Riemannian planar wall which is obtained from a non-Riemannian planar distribution of spinning matter.
\end{abstract}
\newpage
Gravitational collapse of plane symmetric inhomogeneous distribution of matter leading to cosmic walls have been investigated in detail by Holvorcem and Letelier \cite{1}.In this letter we present two examples of metrics which represent cosmic walls obtained from non-Riemannian gravitational collapse.The first is a non-Riemannian wall obtained from a also non-Riemannian planar Taub wall \cite{2} obtained from the General relativistic cosmic wall by a time dependent perturbation.The second is the same time dependent perturbation acting on a Vilenkin Riemannian wall which yields a Riemannian planar thin wall from the gravitational collapse of a non-Riemannian distribution of spin-torsion matter.Let us consider the first metric given by 
\begin{equation}
ds^{2}=\frac{(-dt^{2}+dz^{2})}{(1+kz)^{\frac{1}{2}}}+(1+kz)e^{\frac{{\beta}}{t}}(dx^{2}+dy^{2})
\label{1}
\end{equation}
Where ${\beta}$ is a constant.Note that when time becomes infinity this metric reduces to the well known Taub space-time .Let us now consider the Einstein-Cartan field equation in the quasi-Einsteinian form
\begin{equation}
G_{ij}=T^{EC}_{ij}
\label{2}
\end{equation}
where $T^{EC}_{ij}$ is the energy-stress tensor similar to the Kalb-Hammond energy momentum tensor given by
\begin{equation}
T^{EC}_{ij}=S_{ikl}S_{j}^{kl}-\frac{1}{2}g_{ij}S_{ijk}S^{ijk}
\label{3}
\end{equation}
The Einstein Riemannian tensor componenets $G_{ij}$, where ${i,j=0,1,2,3}$ are given by 
\begin{equation}
G_{tt}={\mu}_{t}^{2}-2{\mu}_{zz}-3{\mu}_{z}^{2}+2{\mu}_{z}{\nu}_{z} 
\label{4}
\end{equation}
and
\begin{equation} 
G_{tz}=2({\mu}_{tz}+{\mu}_{z}{\mu}_{t}-{\mu}_{t}{\nu}_{z})
\label{5}
\end{equation}
which is responsible for the heat flow and
\begin{equation}
G_{zz}=2{\mu}_{tt}+3{\mu}_{t}^{2}-2{\mu}_{z}-{\mu}_{z}^{2}
\label{6}
\end{equation}
which yields the orthogonal pressure $p_{|}$ and finally the component $G_{xx}=G_{yy}$ which is responsible for the parallel pressure $p_{||}$
\begin{equation}
G_{xx}=G_{yy}=e^{2({\mu}-{\nu})}({\mu}_{tt}+{\mu}_{t}^{2}-{\mu}_{zz}-{\mu}_{z}^{2}-{\nu}_{zz})
\label{7}
\end{equation}
for the general plane symmetric space-time
\begin{equation}
ds^{2}=e^{2{\nu}}(-dt^{2}+dz^{2})+e^{2{\mu}}(dx^{2}+dy^{2})
\label{8}
\end{equation}
By choosing the following non-vanishing components of the spin-density tensor $S_{023}=S_{123}$ and by substitution of metric (\ref{1}) into the expressions for the Einstein tensor we obtain the following expressions
\begin{equation}
{\sigma}=S^{2}+(1+kz)^{2}[\frac{2{\beta}}{t^{3}}+\frac{{\beta}^{2}}{t^{4}}]
\label{9}
\end{equation}
where ${\sigma}$ is the surface energy density of the wall and ${S_{123}}^{2}={S_{023}}^{2}=S^{2}$ is the spin-torsion energy density.The remaining equations yield the equations for the heat flow
\begin{equation}
p'_{|}=\frac{{\beta}k(1+kz)}{4t^{2}}+S^{2}
\label{10}
\end{equation}
the orthogonal pressure 
\begin{equation}
p_{|}=S^{2}+(1+kz)(\frac{4{\beta}}{t^{3}}+\frac{3{\beta}^{2}}{4t^{4}})
\label{11}
\end{equation}
and finally a tranversal pressure
\begin{equation} 
p_{||}=S^{2}+(1+kz)e^{\frac{{\beta}}{t}}[\frac{4{\beta}}{t^{3}}+\frac{3{\beta}^{2}}{4t^{4}}]
\label{12}
\end{equation}
One must notice that from these expressions when time goes to infinity the Taub wall collapses to a purely non-Riemannian torsion wall where the surface energy density and the pressures reduces to the spin-torsion density.Notice also that when time approaches the origin these quantities diverge and produce a true singularity in spacetimes with Cartan torsion.Other classes of metrics which decay to cosmic walls may appear elsewhere.
The second example is given by
\begin{equation}
ds^{2}=e^{F}dt^{2}-e^{G}dz^{2}-e^{H}(dx^{2}+dy^{2}) 
\label{13}
\end{equation}
where $H=-4{\pi}{\sigma}(z-t-\frac{{\beta}}{t})$, $(z>0)$ and $F=G=-4{\pi}{\sigma}z$ and we note that when time tends to infinity the metric reduces to the thin cosmic wall metric.Here ${\beta}$ is a constant.Riemannian thin cosmic wall EMT is given by 
\begin{equation} 
T^{i}_{j}={\sigma}diag(1,0,1,1){\delta}(z)
\label{14}
\end{equation}
where ${\sigma}$ represents the matter surface distribution of
the cosmic wall ${\delta}(z)$ is the Dirac delta distribution carachterizing the thin wall.The Einstein-Cartan field equations read  
\begin{equation}
\begin{array}{llll}
T^{EC}_{00}={\frac{1}{32{\pi}}}(H^{2}_{t}-4H_{zz}-3H^{2}_{z}+2H_{z}G_{z})
\nonumber \\
\\
T^{EC}_{01}={\frac{1}{16{\pi}}}(H_{t}G_{z}+H_{t}H_{z})  \nonumber \\
\\
T^{EC}_{11}=\frac{1}{32{\pi}}(-3H^{2}_{t}+H^{2}_{z}-4H_{tt}+2H_{z}G_{z})
\nonumber \\
\\
T^{EC}_{22}=T^{EC}_{33}=\frac{1}{32{\pi}}e^{H-F}(-H^{2}_{t}+H^{2}_{z}-2H_{tt})\nonumber\\
\end{array}
\label{15}
\end{equation} 
here the EMT $T^{EC}_{ij}$ is the total Einstein-Cartan stress-energy tensor given generally by \cite{3}
\begin{equation}
T^{EC}_{ij}=T_{ij}+S_{ikl}S_{j}^{kl}-{\frac{1}{2}}g_{ij}S^{2}
\label{5}
\end{equation}
where $S_{ikl}$ is the torsion tensor and $S^{2}=S_{ijk}S^{ijk}$ is the spin-torsion energy density.This tensor is very similar to the Kalb-Rammond tensor of superstrings.Now we are able to solve the Einstein-Cartan equations for the plane symmetric distribution (\ref{13}).We shall choose the distribution of spin and torsion in such a way that torsion is totally skew-symmetric and the torsion vector is orthogonal to $z=0$ domain wall, that is, the only non-vanishing component of the torsion is $S_{123}$.The $(00)$ equation when time goes to infinity is 
\begin{equation}
T^{EC}_{00}=({\sigma}-S^{2})e^{-4{\pi}{\sigma}z}
\label{16}
\end{equation}
Nevertheless this does not mean that the cosmic wall obtained from the gravitational collapse be non-Riemannian because we shall prove from the other EMT components that torsion vanishes also when time goes to infinity.The component $(01)$ vanishes identically which means that the gravitational collapse does not produce any heat flux.Let us now compute the remaining components $(11)$ and $(22)$.The $(11)$ component yields
\begin{equation}
S^{2}=(S_{123})^{2}=\frac{{\sigma}{\beta}}{4}[\frac{1}{t^{3}}-\frac{{\sigma}{\pi}}{t^{2}}-\frac{3{\pi}{\sigma}{\beta}}{2t^{4}}]e^{-4{\pi}{\sigma}(z-2t-\frac{2{\beta}}{t})}
\label{7}
\end{equation}
we note that when time goes to infinity,torsion vanishes which carachterize the Riemannian nature of the cosmic thin wall obtained after the gravitational collapse.Finally computation of the remaining components of the energy-momentum tensor yields $T^{EC}_{22}=T^{EC}_{33}={\sigma}e^{-4{\pi}{\sigma}z}$ as time goes to infinity, shows to be compatible with the result.This simple example is not meant to produce a general result and a simple examination of this letter shows that a modification of our choice of the metric leads to a possible gravitational collapse of domain walls with heat flux and we can end up with a non-Riemannian space-time.Recently we have investigated \cite{4} the inflating torsion defects produced from dilatonic domain walls and a natural extension of the work discussed here would be to investigate the effect of gravitational collapse of the non-Riemannian walls on inflationary cosmological models.Let us now investigate more realistic models like the non-Riemannian planar thin cosmic wall with heat flow is enough to modify the spin-torsion distribution to trigger the gravitational collapse ,for example by choosing one extra torsion component as $S_{023}$.With this choice the heat flow component $T^{EC}_{01}$ does not vanish anymore and is given by $T^{EC}_{01}=2S_{023}S_{1}^{23}$ which shows that the heat flow is due solely to the spin-torsion distribution.Some years ago H.Soleng \cite{5} have presented a solution of EC field equations representing a cosmic string with heat flow without however discussing the gravitational collapse of non-Riemannian cosmic strings.More general models including non-Riemannian cosmic walls from gravitational collapse of non-Riemannian space-times can be obtained by simply modifying the nature of space-time metric (\ref{1}).Work in this direction is now under progress.	  
\section*{Acknowledgments}
\paragraph*{}
I am very much indebt to Professors P.S.Letelier,Harald Soleng ,P.Steinhardt and P.Shellard for enlightening discussions on the subject of this paper.Financial supports from CNPq. (Brazilian Government Agency) and Universidade do Estado do Rio de Janeiro (UERJ) are gratefully acknowledged.Financial support and hospitality of the Isaac Newton Institute for Mathematical Sciences of Cambridge University,where part of this work was carried out is also acknowledged. 

\end{document}